\documentclass[reprint,superscriptaddress,amsmath,amssymb,aps,prl]{revtex4-2}

\usepackage{graphicx}
\usepackage{dcolumn}
\usepackage{bm}
\usepackage{physics}
\usepackage{xcolor}
\usepackage[colorlinks=true, allcolors=blue]{hyperref}
\definecolor{mag}{RGB}{255,0,255}

\begin{document}

\title{Theory of Nb-Zr alloy superconductivity and first experimental demonstration for superconducting radio-frequency cavity applications}

\author{Nathan S. Sitaraman} 
\thanks{These authors contributed equally.}
\affiliation{
 Department of Physics,
 Cornell University, Ithaca, New York 14853, USA
}

\author{Zeming Sun}
\thanks{These authors contributed equally.}
\affiliation{
 Cornell Laboratory for Accelerator-Based Sciences and Education, Cornell University, Ithaca, New York 14853, USA
}
\author{Ben Francis}
\thanks{These authors contributed equally.}
\affiliation{
 Department of Physics and Astronomy,
 Brigham Young University, Provo, Utah 84602, USA
}
\author{Ajinkya C. Hire}
\thanks{These authors contributed equally.}
\affiliation{
 Quantum Theory Project, University of Florida, Gainesville, Florida 32611, USA
}
\author{Thomas Oseroff}
\affiliation{
 Cornell Laboratory for Accelerator-Based Sciences and Education, Cornell University, Ithaca, New York 14853, USA
}

\author{Zhaslan Baraissov}
\affiliation{
 Department of Applied Physics,
 Cornell University, Ithaca, New York 14853, USA
}

\author {Tom\'as A. Arias}
\email{taa2@cornell.edu}
\affiliation{
 Department of Physics,
 Cornell University, Ithaca, New York 14853, USA
}

\author{Richard Hennig}
\email{rhennig@ufl.edu}
\affiliation{
 Quantum Theory Project, University of Florida, Gainesville, Florida 32611, USA
}

\author{Matthias U. Liepe}
\email{mul2@cornell.edu}
\affiliation{
 Cornell Laboratory for Accelerator-Based Sciences and Education, Cornell University, Ithaca, New York 14853, USA
}

\author{David A. Muller}
\affiliation{
 Department of Applied Physics,
 Cornell University, Ithaca, New York 14853, USA
}

\author{Mark K. Transtrum}
\email{mktranstrum@byu.edu}
\affiliation{
 Department of Physics and Astronomy,
 Brigham Young University, Provo, Utah 84602, USA
}

\collaboration{Center for Bright Beams}

\date{\today}

\begin{abstract}
Niobium-zirconium (Nb-Zr) alloy is an old superconductor that is a promising new candidate for superconducting radio-frequency (SRF) cavity applications. Using density-functional and Eliashberg theories, we show that addition of Zr to a Nb surface in small concentrations increases the critical temperature $T_c$ and improves other superconducting properties.
Furthermore, we calculate $T_c$ for Nb-Zr alloys across a broad range of Zr concentrations, showing good agreement with the literature for disordered alloys as well as the potential for significantly higher $T_c$ in ordered alloys near 75\%Nb/25\%Zr composition. We provide experimental verification on Nb-Zr alloy samples and SRF sample test cavities prepared with either physical vapor or our novel electrochemical deposition recipes. These samples have the highest measured $T_c$ of any Nb-Zr superconductor to date and indicate a reduction in BCS resistance compared to the conventional Nb reference sample; they represent the first steps along a new pathway to greatly enhanced SRF performance. Finally, we use Ginzburg-Landau theory to show that the addition of Zr to a Nb surface increases the superheating field $B_{sh}$, a key figure of merit for SRF which determines the maximum accelerating gradient at which cavities can operate.
\end{abstract}

\maketitle

\paragraph{Introduction.}

Superconducting radio-frequency (SRF) cavities are a key component of particle accelerators used, e.g., as light sources \cite{SunRef4} and colliders \cite{SunRef5}.  These cavities are extremely reliant on the superconducting properties of the cavity surface.  Of particular interest are the RF resistance, which determines energy dissipation and cooling costs, and the superheating field, which determines the maximum accelerating gradient and thus the number of cavities required \cite{padamsee1993physics,padamsee2001science}. Further optimization of these quantities is necessary to enable more energy-efficient and compact accelerators and requires the exploration of superconductors beyond industry-standard niobium (Nb).

Many efforts have explored adding other elements to the SRF surface of Nb cavities, such as tin \cite{posen2015proof,posen2017nb3sn,posen2021advances} and nitrogen \cite{Burton}.  One of the limitations of these materials is the difficulty of producing uniform layers of the necessary phases and compositions that achieve the desired improvements in superconducting properties.  A possible material that avoids these complications is niobium-zirconium (Nb-Zr).  Nb-Zr alloys retain a body-centered cubic (bcc) crystal structure, the same as Nb, across a broad compositional range up to about 50\% Zr \cite{SunRef6}, and form a stable, insulating Zr native oxide which should cause little dissipation under RF conditions.

The superconducting properties of Nb-Zr alloys, however, are not as well characterized as those of other Nb-based superconductors. Literature studies \cite{SunRef1,hulm1961} have focused on bulk random alloy samples very different from the thin-film alloy layers necessary for SRF. Furthermore, these bulk random-alloy samples showed a relatively modest enhancement of the critical temperature $T_c$ compared to elemental Nb, far less than would be expected based on the rigid band shift theory proposed by early researchers \cite{hulm1961}.

As for the superheating field $B_{sh}$, previous studies have addressed homogeneous superconductors \cite{chapman1995superheating,dolgert1996superheating,transtrum2011superheating,catelani2008temperature,lin2012effect} and layered superconductors \cite{liarte2016ginzburg,kubo2014radio,gurevich2015maximum,posen2015shielding,kubo2016multilayer}, but efforts to improve SRF cavity performance have revealed that material inhomogeneities are key factors which affect $B_{sh}$ \cite{padamsee2001science}.
Recent studies of $B_{sh}$ for inhomogeneous superconductors include materials with continuously-varying impurity concentrations \cite{ngampruetikorn2019effect} and the effect of spatial variations in $T_c$ \cite{pack2020vortex}.

To better understand the Nb-Zr alloy system, we use density-functional theory (DFT) and Eliashberg theory in the dilute limit ($\leq$25\% Zr), where the virtual crystal approximation (VCA) allows us to calculate a variety of superconducting properties.
We confirm that the addition of Zr to Nb increases $T_c$.

We then switch from VCA to a supercell approach in order to directly account for lattice relaxation effects, which become important at higher Zr concentrations (up to 50\% Zr), and focus on $T_c$. With this approach, we find good agreement with the literature for random alloys at all concentrations, and we investigate the potential for ordered alloy structures to surpass the limitations of the random alloy.

We synthesize Nb-Zr alloy surface layers of different Zr doping profiles using physical vapor and electrochemical methods and examine their material and RF superconducting properties. We find that the measured $T_c$ meets or exceeds literature values for Nb-Zr random alloys, in agreement with our calculations. We show that the enhanced $T_c$ likely translates to improved performance under RF conditions by comparing the BCS resistance of the alloyed sample to that of a reference Nb sample.

Finally, we use GL theory to investigate how Nb-Zr surface layers with different thicknesses and compositions affect $B_{sh}$.  We incorporate spatial variations of multiple superconducting properties using realistic profiles of Zr concentration.  We find that addition of Zr to Nb noticeably improves $B_{sh}$, making Nb-Zr an excellent candidate for SRF applications.

\paragraph{Superconducting properties of Nb-Zr in the dilute limit.}

We calculate the superconducting properties of Nb-Zr alloys in the dilute limit using VCA. We use Quantum Espresso (QE) \cite{qe1,qe2,qe3} to perform the DFT calculations. We use the Perdew–Burke-Ernzerhof \cite{PBE} functional for the exchange-correlation energy and norm conserving pseudopotentials \cite{Hamann2013,Schlipf2015}. We use the Methfessel-Paxton smearing scheme \cite{Methfessel1989} to smear the electrons during lattice relaxation. For the electron-phonon calculations, we use the tetrahedron method \cite{PhysRevB.89.094515} as implemented in QE, with a $\vb{q}$-mesh of $8 \times 8 \times 8$ and a $\vb{k}$-mesh of $24 \times 24 \times 24$ to calculate the Eliashberg function $\alpha^2F$. Finally, we calculate the superconducting gap using isotropic Eliashberg theory, as implemented in the EPW code \cite{epw1,epw2}.

\begin{figure}[htbp]
    \centering
    \includegraphics[width=\linewidth]{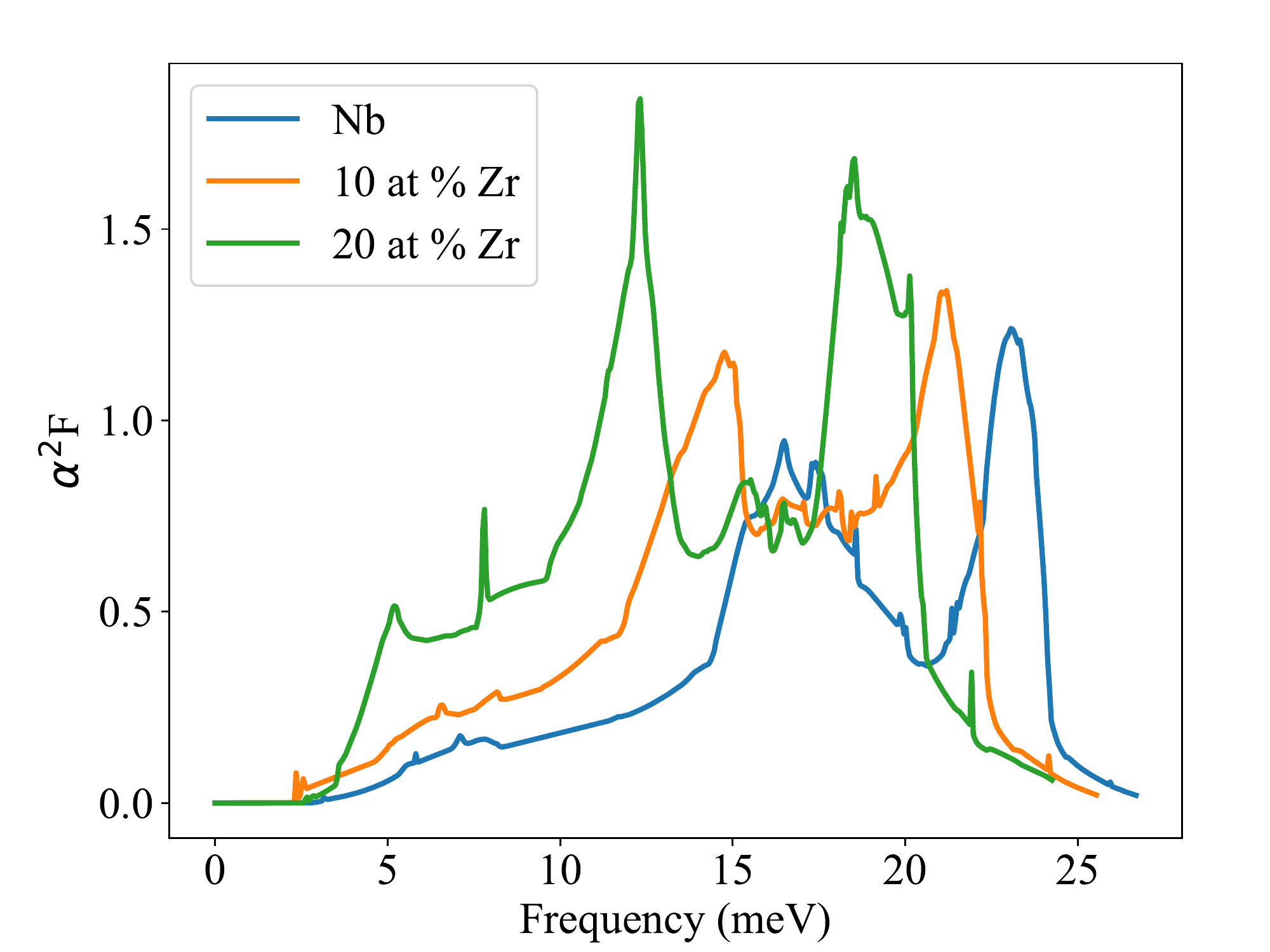}
    \caption{Calculated $\alpha^2F$ of Nb-Zr alloys. As Zr concentration increases, phonons at lower frequency couple strongly with electrons, leading to an increase in the electron-phonon coupling constant.}
    \label{fig:figure1}
\end{figure}

Table \ref{tab:table1} shows the calculated superconducting properties for Zr concentrations ranging from 0 to 25\%. The calculated $T_c$ agree well between the three methods used to calculate the critical temperature - Allen-Dynes \cite{allen-dynes}, Xie \cite{Xie_2022}, and isotropic Eliashberg. As Zr concentration increases, so do both the electron-phonon coupling constant and the $T_c$ of the alloy. This increase in $T_c$ can be attributed to softening of the phonons, as shown in Fig.~\ref{fig:figure1}, and an increase in the density of states at the Fermi level $N(0)$. 
\begin{table*}
    \centering
    \caption{Calculated superconducting properties of Nb-Zr alloys using VCA and GL theory, including the electron-phonon coupling constant $\lambda^e$ and the superconducting gap $\Delta$. $T_c$ was calculated using the Allen-Dynes equation, isotropic Eliashberg theory, and Xie22\cite{Xie_2022} using $\mu^*=0.16$. $\lambda_L$ and $\xi$ have been rescaled to match experimental values for Nb.}
    \label{tab:table1}
    \begin{ruledtabular}
    \begin{tabular}{lccccccccccc}
        Composition & $\lambda^e$ & $\Delta$(4K) (meV) & $v_F$ (m/s) & $N(0)$ (states/eV/\AA$^3$) & $T_c^{\text{AD}}$ (K) & $T_c^{\text{El}}$ (K) & $T_c^{\text{Xie}}$ (K) & $\lambda_L$ (nm) & $\xi$ (nm) & $B_c$ (mT) & $\kappa$ \\
        \hline
        Nb                      & 1.13 & 2.4 & $8.50 \times 10^5$ & 0.082 & 11.24 & 14.2 & 10.39 & 39.0 & 38.0 & 157 & 1.03 \\
        Nb$_{0.95}$Zr$_{0.05}$  & 1.30 & 3.0 & $8.20 \times 10^5$ & 0.086 & 13.60 & 16.8 & 13.46 & 39.5 & 30.3 & 195 & 1.30 \\
        Nb$_{0.90}$Zr$_{0.10}$  & 1.69 & 4.0 & $8.09 \times 10^5$ & 0.091 & 17.18 & 20.8 & 18.3 & 38.9 & 23.7 & 253 & 1.64 \\
        Nb$_{0.80}$Zr$_{0.20}$  & 2.37 & 5.3 & $7.79 \times 10^5$ & 0.103 & 21.17 & 25.3 & 24.28 & 38.0 & 18.5 & 331 & 2.05 \\
        Nb$_{0.75}$Zr$_{0.25}$  & 2.85 & 5.8 & $7.70 \times 10^5$ & 0.106 & 22.02 & 26.7 & 26.49 & 37.9 & 17.6 & 350 & 2.16 \\
    \end{tabular}
    \end{ruledtabular}
\end{table*}

\paragraph{Superconducting properties of Nb-Zr alloys at higher Zr concentrations.}

\begin{figure}[htbp]
    \centering
\includegraphics[width=0.4\textwidth]{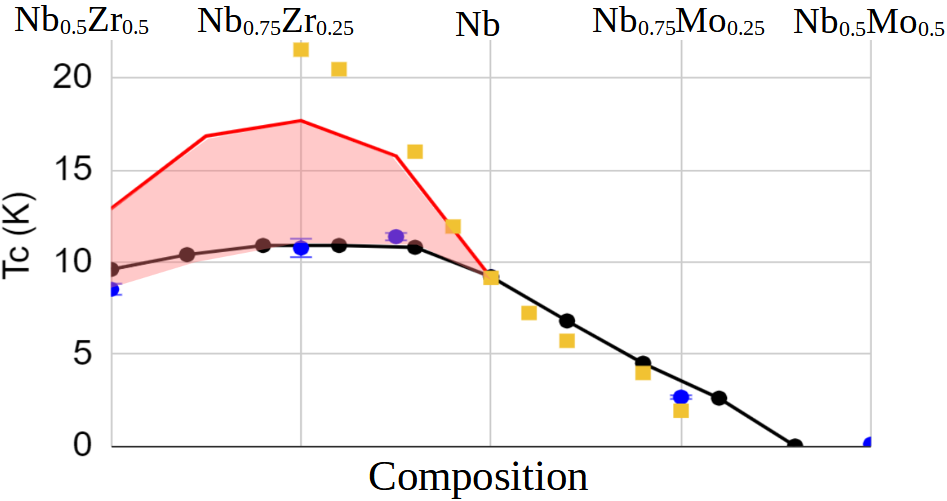}
    \caption{Alloy $T_c$ vs.~composition for experiment (black, \cite{hulm1961}), VCA (yellow squares), random supercell theory (blue circles), and Jahn-Teller-Peierls stability limit (red).
        }
    \label{fig:figure2}
\end{figure}

Previous experimental studies of Nb$_{1-n}$Zr$_n$ bcc random alloys have found moderately increased $T_c$ and greatly increased upper critical field $B_{c2}$ compared to Nb across a broad compositional range from $n=0$ to $n=0.5$ \cite{hulm1961,SunRef1}. Alloys made with molybdenum (Mo), Nb's other 5th row neighbor, have $T_c$'s that drop off quickly with increasing Mo concentration. This data (Fig.~\ref{fig:figure2}`experiment' curve) is in qualitative agreement with the simple alloy theory originally used to describe bcc alloy superconductivity, namely that there is a universal d-band shape, and alloying simply shifts the position of the Fermi level by adding or removing d-band electrons, thus altering $N(0)$ and therefore $T_c$ \cite{hulm1961,BCS}. Quantitatively, this effect is captured in our VCA calculations (Fig.~\ref{fig:figure2} 'VCA' curve), which show good agreement with experiment for Nb-Mo, but significantly overestimate $T_c$ for Nb-Zr. This indicates that, while VCA may accurately describe Nb-Mo and dilute Nb-Zr alloys, it misses important nonlinear effects in the Nb-Zr system that alter the d-band shape and tend to lower $T_c$.

Exploration of the full compositional range of Nb$_{1-n}$Zr$_n$ therefore requires an analysis of Nb-Zr which accounts fully for the effects of Zr substitutions in the Nb lattice; for this we employ the supercell method. Specifically, for each $n$, we construct 48-atom supercells, substitute the required number of Nb atoms with Zr at random sites, and fully relax the structure, using the JDFTx software package with the PBE exchange-correlation functional and ultrasoft pseudopotentials \cite{Sundararaman,PBE,Psp}. Averaging the resulting density of states (DOS) over multiple configurations at each composition yields an expected macroscopic $N(0)$. We then estimate $T_c$ as a function of composition through the BCS $N(0)$-$T_c$ relationship, resulting in the `random alloy' predictions displayed in Fig.~\ref{fig:figure2}. We find quite good agreement with the random-alloy experiments for both Nb-Mo and Nb-Zr, indicating that our calculated $N(0)$--composition relationship correctly captures the important nonlinear effects that determine $T_c$ in Nb-Zr random alloys.

\begin{figure}[htbp]
    \centering
    \includegraphics[width=0.45\textwidth]{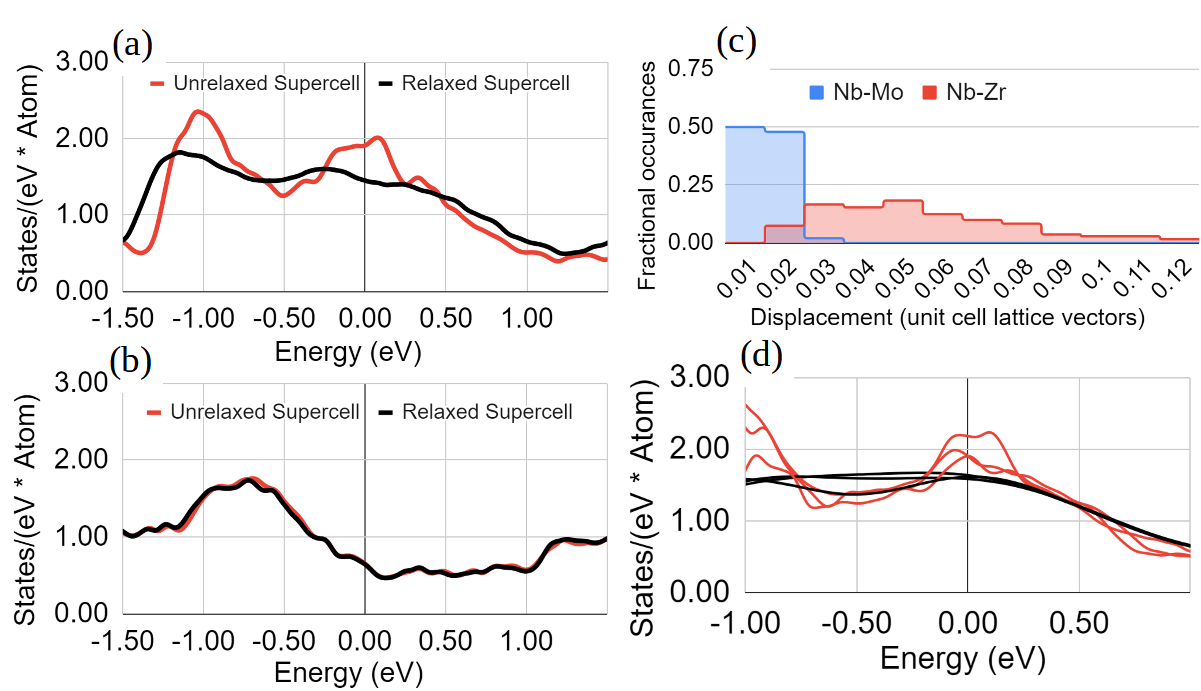}
    \caption{
        Electronic DOS for 50/50 Nb-Zr (a) and Nb-Mo (b) random alloys before atomic relaxation (red) and after atomic relaxation (black). (c) Histogram of atomic displacements in 50/50 Nb-Zr (red) and Nb-Mo (blue) alloys. (d) DOS curves for three different ordered 50/50 Nb-Zr alloys before (red) and after (black) applying electronic smearing to reduce $N(0)$ and stabilize the structure. 
        }
    \label{fig:figure3}
\end{figure}

Examining our supercell calculations, we find that, for atoms at their unrelaxed, ideal bcc locations, the electronic DOS for Nb-Zr alloys resemble that of Nb with a rigid band shift, similar to what would be found in VCA. However, we find that the atoms tend to move away from their ideal bcc locations in a way that significantly reduces $N(0)$. This final, reduced $N(0)$ value varies only modestly with composition up to 50\% Zr, explaining the modest variations that are observed in $T_c$.  Figure~\ref{fig:figure3}a illustrates this relaxation effect for the 50/50 Nb-Zr random alloy, where the initially large DOS peak near the Fermi level essentially vanishes. Moreover, consistent with our observation that simple alloy theories work well for Nb-Mo, Fig.~\ref{fig:figure3}b shows this relaxation effect to be absent in the Nb-Mo system.

Fundamentally, the Nb-Zr alloy minimizes its energy by lowering the energies of occupied electronic states near the Fermi level, resulting in a reduced $N(0)$. This relaxation does not occur in Nb-Mo, which has fewer states near the Fermi level, and thus less ability to lower its energy in this manner. The redistribution of electronic energy levels in Nb-Zr is accomplished by the relaxation of lattice degrees of freedom, i.e., by the movement of atoms away from their ideal lattice positions. This brings us to a key difference between random Nb-Zr alloys, which have been our focus so far, and ordered Nb-Zr alloys. It has been pointed out that the symmetries of highly ordered structures can effectively limit the degrees of freedom for lattice relaxation \cite{allen-dynes}. In our case, this could result in a crystal with electronic structure similar to the VCA or the unrelaxed-random-alloy electronic structure and thus a much higher $T_c$. It may be possible to grow such an ordered alloy in a low-temperature thin-film process such as those often employed in SRF cavity production.

We estimate the maximum attainable $T_c$ of an ordered Nb-Zr alloy of a given composition as follows. We note that the same force which drives ionic displacements in the Nb-Zr random alloy tends to break the symmetries of ordered Nb-Zr structures
. Indeed, this Jahn-Teller-Peierls effect has been proposed as a driving force for bcc phase instability in binary alloys, including Nb-Zr \cite{widom2018}. The magnitude of this driving force is directly related to the $N(0)$ of the unrelaxed structure, which we can tune artificially by broadening the electronic states in our calculations. We then determine the limiting value of $N(0)$ beyond which the Jahn-Teller-Peierls instability causes the structure to break its symmetry and relax to a lower-symmetry state with reduced $N(0)$. Fig.~\ref{fig:figure3}d, for example, shows the result of this process for three high symmetry structures at the same Zr concentration of $n=0.5$, showing that they all exhibit very similar $N(0)$ just before the onset of the instability. Finally, this stability-limited Fermi-level DOS can be used to determine the limiting $T_c$ as a function of composition through BCS theory. Figure~\ref{fig:figure2} (`Jahn-Teller-Peierls stability limit' curve) displays the results of the above process. We generally expect that the $T_c$ values for Nb$_{1-n}$Zr$_n$ will all lie within the pink shaded region, between the random alloy result and the limiting Jahn-Teller-Peierls value for any special, ordered structure. For example, an Eliashberg calculation for a specific ordered structure at $n=0.25$ gives an electron-phonon coupling constant $\lambda^e = 1.93$ compared to a value of $\lambda^e = 1.18$ for Nb. To estimate the corresponding $T_c$ enhancement, we use the Allen-Dynes formula with $\mu^* = 0.1$ and an additive shift in the $\lambda^e$ parameter to account for spin-fluctuations and match the experimental $T_c$ of bulk Nb \cite{allen-dynes,Bekaert}. This yields a $T_c$ of 15.7~K for the ordered $n=0.25$ alloy, close to our predicted stability limit of 17.7~K. This prediction represents an additional 50\% enhancement over the random alloy $T_c$, similar to what we indeed observe in our experiments (see below).

\paragraph{Experimental verification.}
In order to validate our theoretical predictions, we measure the superconducting properties and RF performance of different Nb-Zr surface profiles. Samples were prepared using e-beam evaporation of a Zr target (base pressure: $1.3 \times 10^{-6}$ torr) on the Nb surface or electrochemical reaction with the Nb surface, followed by thermal annealing under $2 \times 10^{-7}$ torr vacuum, and a subsequent HF etch. The initial film thickness (20 -- 40 nm) and post annealing conditions (600 -- 1000 \textdegree C for 1/3 -- 10 h) were varied to modify the surface Zr atomic concentration. 

\begin{figure}[htbp]
    \centering
    \includegraphics{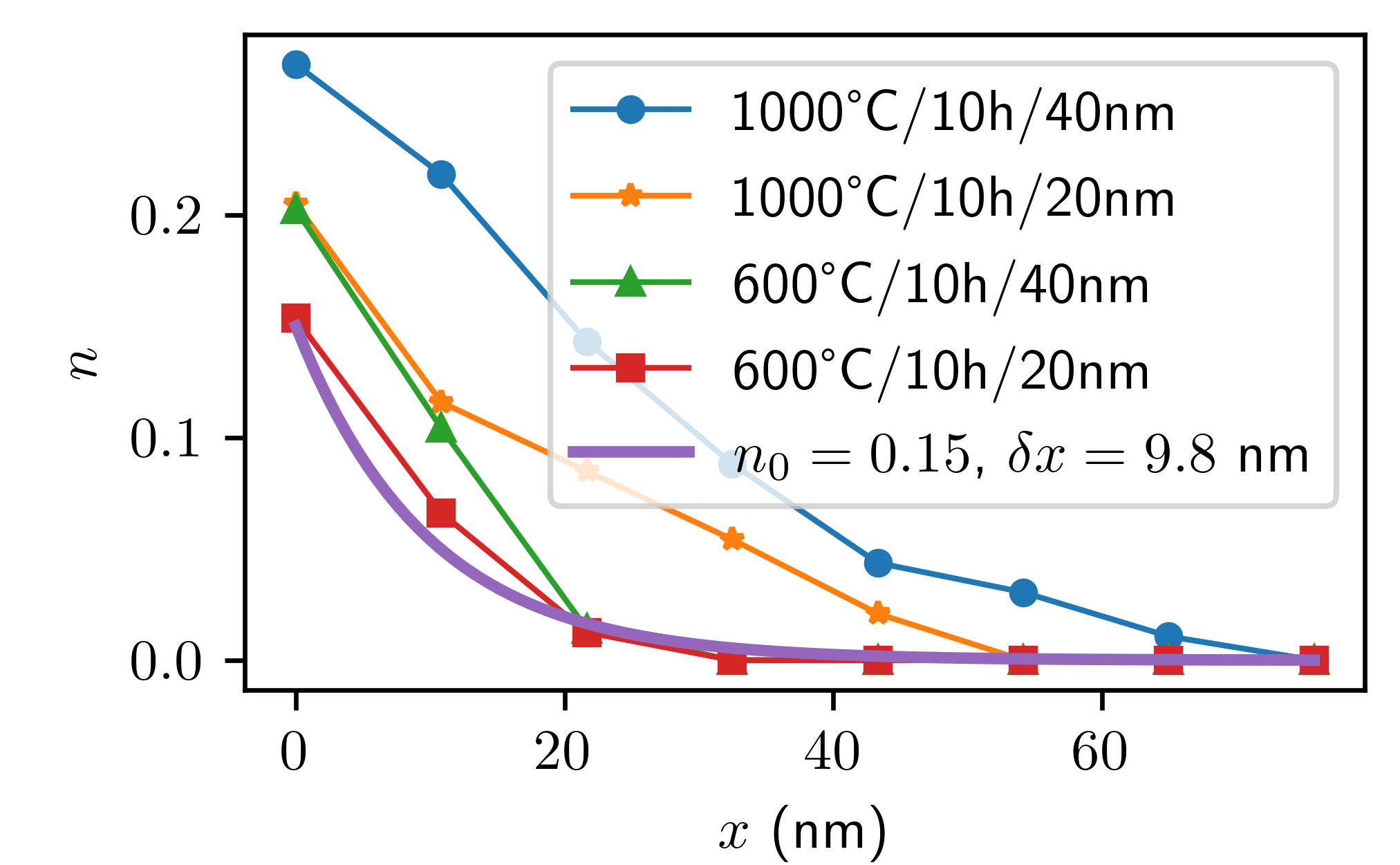}
    \caption{
        Zr concentration $n$ vs.~depth $x$ for Nb-Zr samples prepared under various annealing conditions, as well as for the profile $n(x)$ in Eq.~\eqref{eq:equation4}.
        }
    \label{fig:figure4}
\end{figure}

\begin{figure*}
    \centering
    \includegraphics[width=6in]{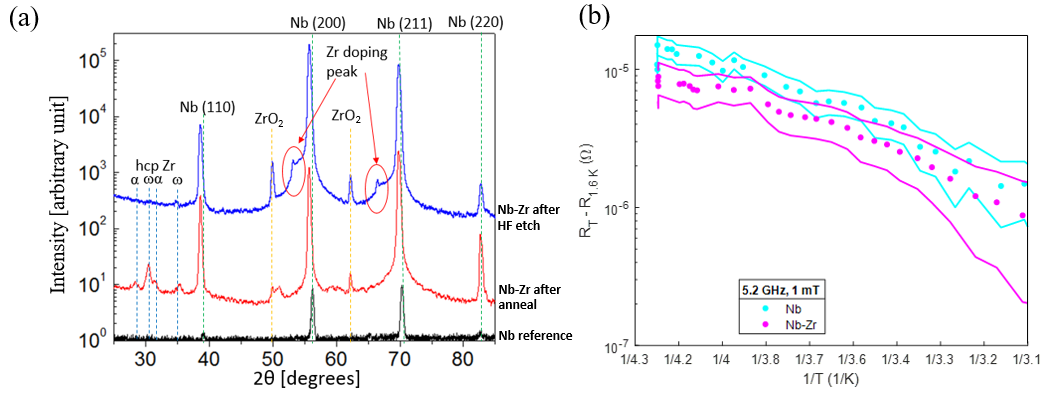}
    \caption{
        Structural and superconducting properties of Nb-Zr samples. (a) X-ray diffraction pattern for evaporation-based samples annealed at 600\textdegree C for 10 h, with a subsequent HF etch. (b) BCS resistance before and after Nb-Zr alloying.   
        }
    \label{fig:figure5}
\end{figure*}

As probed by X-ray photoelectron spectroscopy, we observe 15--27 at.\% Zr at the surface of evaporation-based samples (Fig.~\ref{fig:figure4}) as well as significant oxygen concentrations.  From X-ray diffraction (Fig.~\ref{fig:figure5}a), we infer that substitutional Zr doping was achieved, as evidenced by the doping peaks at lower diffraction angles compared to a Nb cubic reference. Zr's and Nb's bcc lattice parameters are 0.354 nm and 0.330 nm, respectively \cite{SunRef8}. Based on Vegard’s law, these doping peaks are induced by lattice enlargement when Zr dopants are incorporated into the cubic structure.  Moreover, the HF etch entirely eliminates the hexagonal Zr phases that appeared on the outermost region after annealing. Note that the hexagonal $\alpha$ and $\omega$ phases have low $T_c$'s of 7~K and 4~K, respectively \cite{SunRef1,SunRef7}, so avoiding these hexagonal phases is critical to obtaining high $T_c$. Additionally, the only oxide we detect is ZrO$_2$, which is ideal for SRF applications due to its wide bandgap. 

Resistance measurements using a Physical Property Measurement System under the AC transport mode demonstrate a $T_c$ of up to 13.5~K for evaporation-based samples that were annealed under 600 \textdegree C for 10 h. Even more impressively, the flux expulsion measurements indicate our electrochemically fabricated samples have a higher $T_c$ of 16~K. We point out that the 13.5~K and 16~K $T_c$ values are significantly higher than the literature-reported 11~K $T_c$ for Nb-Zr bulk alloys \cite{SunRef1} and the 7.4~K $T_c$ measured in sputtered Nb-Zr thin films \cite{SunRef2}. One can infer that retention of an ordered cubic structure is critical to achieving a higher $T_c$ than random alloys, which matches well with our DFT simulations.

To assess the use of Nb-Zr alloys for SRF accelerator applications, we employ an electrochemical method to scale up the alloying process to be compatible with the Cornell sample test cavity \cite{SunRef3}. Indeed, as shown in Fig.~\ref{fig:figure5}b, after Nb-Zr alloying, the 5.2 GHz low-field BCS surface resistance is trending lower, which is consistent with the expected benefit of the high $T_c$ of the Nb-Zr material. This first RF demonstration of Nb-Zr alloys establishes a new direction for SRF cavities with high $T_c$ and low surface resistance.
As we now show, these cavities also have the potential for high superheating fields.

\paragraph{GL predictions of \texorpdfstring{$B_{sh}$}{Bsh}.}
To calculate $B_{sh}$, we employ GL theory to simulate Nb-Zr alloys in cavity surface layers using the data from Table \ref{tab:table1}.
The main object of the theory is the free energy density $\mathcal{F}$ of the system \cite{tinkham2004introduction,kopnin2001theory,de2018superconductivity}, which includes both an expansion in a superconducting order parameter $\psi$ (with phenomenological coefficients $\alpha$, $\beta$, and $\gamma$) and a contribution from the total magnetic field $\vb{B} = \vb{B}_a - \nabla\times \vb{A}$, where $\vb{B}_a$ is the applied field and $\vb{A}$ is the magnetic vector potential:
\begin{equation}
    \mathcal{F} = \alpha\lvert\psi\rvert^2 + \frac{\beta}{2}\lvert\psi\rvert^4 + \gamma\abs{\left(-i\hbar\nabla - 2e\vb{A}\right)\psi}^2 + \frac{B^2}{2\mu_0}.
\end{equation}
$B_{sh}$ is the value of $\vb{B}_a$ at which the superconducting Meissner state becomes unstable.

It can be shown that $\alpha$, $\beta$, and $\gamma$ are related to the London penetration depth $\lambda_L$, coherence length $\xi$, and thermodynamic critical field $B_c$:
\begin{align}
    \label{eq:equation2}
    \lambda_L^2 &= \frac{\beta}{8\mu_0 e^2\abs{\alpha}\gamma}, &
    \xi^2 &= \hbar^2\frac{\gamma}{\abs{\alpha}}, &
    B_c &= \abs{\alpha}\sqrt{\frac{\mu_0}{\beta}}.
\end{align}
We use Eq.~\eqref{eq:equation2} to calculate $\lambda_L$, $\xi$, and $B_c$ for each of the compositions in Table \ref{tab:table1}.
We observe a favorable increase in $B_c$ as Zr concentration increases.

We calculate the GL parameter $\kappa \equiv \lambda_L/\xi$ for each of the compositions in Table \ref{tab:table1} and note that addition of Zr raises $\kappa$.
At $\kappa_c \approx 1.15$, the type of instability which disrupts the superconducting state changes \cite{transtrum2011superheating}.
Destabilizing perturbations can be decomposed into Fourier modes with wave number $k$.
Below $\kappa_c$, the instability is due to the $k = 0$ mode (i.e., uniform penetration of magnetic flux).
Above $\kappa_c$, the instability results from a mode with $k > 0$.
When calculating $B_{sh}$ for Nb-Zr alloys, we are careful to search across perturbations with $k > 0$ to look for this transition, so as not to overestimate $B_{sh}$.

We simulate inhomogeneous superconductors by treating the coefficients $\alpha$, $\beta$, and $\gamma$ as local quantities that vary spatially with material composition.
For clean superconductors, these coefficients have been shown \cite{kopnin2001theory,de2018superconductivity} to depend on $N(0)$, $T_c$, and the Fermi velocity $v_F$:
\begin{align}
    \label{eq:equation3}
    \alpha &= \frac{N(0)}{T_c}(T - T_c), &
    \beta  &= \frac{7\zeta(3)N(0)}{8\pi^2 k_B^2 T_c^2}, &
    \gamma &= \beta\frac{v_F^2}{6}.
\end{align}
We use these relationships to obtain $\alpha(x)$, $\beta(x)$, and $\gamma(x)$ as follows.
First, we consider a concentration profile $n(x)$ in which Zr concentration $n$ falls off exponentially with depth $x$ in the material from a surface value $n_0$ on a scale set by $\delta x$:
\begin{equation}
    \label{eq:equation4}
    n(x) = n_0 e^{-x/\delta x}.
\end{equation}
(see Fig.~\ref{fig:figure4}).
We then interpolate the values of $N(0)$, $T_c^\text{AD}$, and $v_F$ in Table \ref{tab:table1} for intermediate concentrations $n$ to obtain $N(0;x)$, $T_c(x)$, and $v_F(x)$.
Finally, we compose these with Eq.~\eqref{eq:equation3} to obtain $\alpha(x)$, $\beta(x)$, and $\gamma(x)$.

\begin{figure}[htbp]
    \centering
    \includegraphics{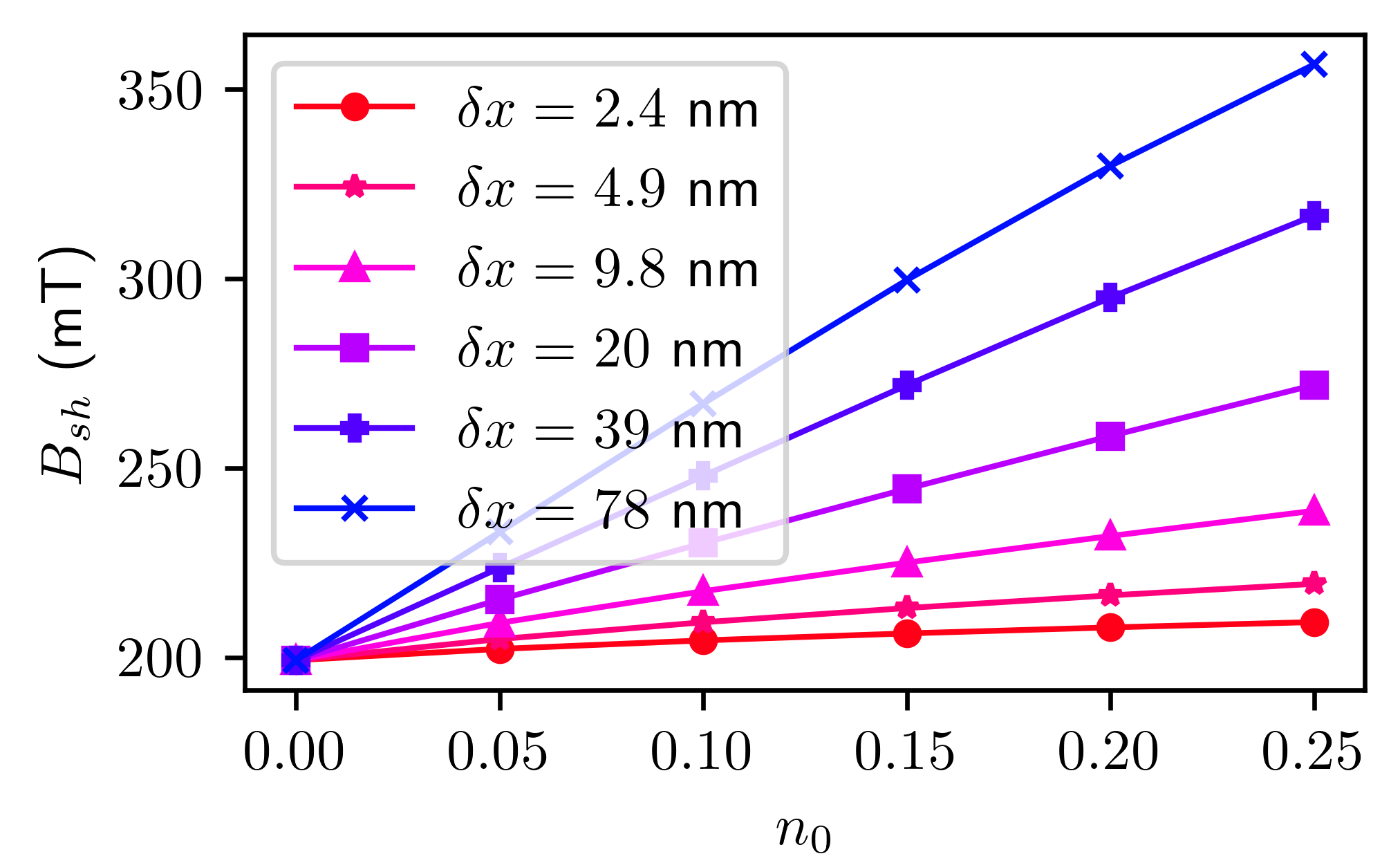}
    \caption{
        Calculated $B_{sh}$ of Nb-Zr surface layers with varying compositions [see Eq.~\eqref{eq:equation4}], rescaled by the value of $B_c$ for Nb in Table \ref{tab:table1}.
        }
    \label{fig:figure6}
\end{figure}

We calculate $B_{sh}$ using the linear stability analysis described in \cite{transtrum2011superheating} for various choices of $n_0$ and $\delta x$ (Fig.~\ref{fig:figure6}).
In all cases, $B_{sh}$ increases as the total Zr content is increased, consistent with the improvements in other superconducting properties listed in Table \ref{tab:table1}.

\paragraph{Conclusion.}

In summary, we have proposed and demonstrated a new type of SRF surface via Nb-Zr alloying that enables:
(1) a high $T_c$ of 16~K that minimizes energy dissipation and cryogenic costs;
(2) a promising predicted enhancement of $B_{sh}$ up to 350 mT that allows approximately 85 MV/m accelerating gradients.
We find an excellent match between theoretical and experimental results, which show improvement of $T_c$ after Zr is incorporated into the Nb lattice, and our RF results indicate a reduction of BCS resistance. Moreover, our theoretical predictions provide a viable road map to further tune and improve $T_c$ and $B_{sh}$. Overall, we have demonstrated that Nb-Zr alloys promise to be a new, feasible technology for accelerator physics. 

\begin{acknowledgments}
\paragraph{Acknowledgments.}
A.H.~and R.H.~performed the VCA calculations.
N.S.S.~and T.A.A.~performed the supercell alloy calculations and thank Michelle Kelley and Ravishankar Sundararaman for helpful conversations.
Z.S., T.O., Z.B., D.A.M., and M.U.L.~completed the experimental verification, including material growth, characterization, and RF performance evaluation.
Z.S.~acknowledges Dr.~Darrah K.~Dare and Katrina Howard for experimental assistance.
B.L.F.~and M.K.T.~did the GL predictions of $H_{sh}$.
All authors revised the manuscript.
This work made use of the Cornell Center for Materials Research Shared Facilities which are supported through the NSF MRSEC program (DMR-1719875) and was performed in part at the Cornell NanoScale Facility, an NNCI member supported by NSF Grant NNCI-2025233.
This work was supported by the U.S.~National Science Foundation under Award PHY-1549132, the Center for Bright Beams.
\end{acknowledgments}


\begin{thebibliography}{47}%
\makeatletter
\providecommand \@ifxundefined [1]{%
 \@ifx{#1\undefined}
}%
\providecommand \@ifnum [1]{%
 \ifnum #1\expandafter \@firstoftwo
 \else \expandafter \@secondoftwo
 \fi
}%
\providecommand \@ifx [1]{%
 \ifx #1\expandafter \@firstoftwo
 \else \expandafter \@secondoftwo
 \fi
}%
\providecommand \natexlab [1]{#1}%
\providecommand \enquote  [1]{``#1''}%
\providecommand \bibnamefont  [1]{#1}%
\providecommand \bibfnamefont [1]{#1}%
\providecommand \citenamefont [1]{#1}%
\providecommand \href@noop [0]{\@secondoftwo}%
\providecommand \href [0]{\begingroup \@sanitize@url \@href}%
\providecommand \@href[1]{\@@startlink{#1}\@@href}%
\providecommand \@@href[1]{\endgroup#1\@@endlink}%
\providecommand \@sanitize@url [0]{\catcode `\\12\catcode `\$12\catcode
  `\&12\catcode `\#12\catcode `\^12\catcode `\_12\catcode `\%12\relax}%
\providecommand \@@startlink[1]{}%
\providecommand \@@endlink[0]{}%
\providecommand \url  [0]{\begingroup\@sanitize@url \@url }%
\providecommand \@url [1]{\endgroup\@href {#1}{\urlprefix }}%
\providecommand \urlprefix  [0]{URL }%
\providecommand \Eprint [0]{\href }%
\providecommand \doibase [0]{https://doi.org/}%
\providecommand \selectlanguage [0]{\@gobble}%
\providecommand \bibinfo  [0]{\@secondoftwo}%
\providecommand \bibfield  [0]{\@secondoftwo}%
\providecommand \translation [1]{[#1]}%
\providecommand \BibitemOpen [0]{}%
\providecommand \bibitemStop [0]{}%
\providecommand \bibitemNoStop [0]{.\EOS\space}%
\providecommand \EOS [0]{\spacefactor3000\relax}%
\providecommand \BibitemShut  [1]{\csname bibitem#1\endcsname}%
\let\auto@bib@innerbib\@empty
\bibitem [{\citenamefont {Decking}\ \emph {et~al.}(2020)\citenamefont {Decking}
  \emph {et~al.}}]{SunRef4}%
  \BibitemOpen
  \bibfield  {author} {\bibinfo {author} {\bibfnamefont {W.}~\bibnamefont
  {Decking}} \emph {et~al.},\ }\bibfield  {title} {\bibinfo {title} {A
  {MHz}-repetition-rate hard {X-ray} free-electron laser driven by a
  superconducting linear accelerator},\ }\href
  {https://doi.org/10.1038/s41566-020-0607-z} {\bibfield  {journal} {\bibinfo
  {journal} {Nature Photonics}\ }\textbf {\bibinfo {volume} {14}},\ \bibinfo
  {pages} {391} (\bibinfo {year} {2020})}\BibitemShut {NoStop}%
\bibitem [{\citenamefont {Sicking}\ and\ \citenamefont
  {Strom}(2020)}]{SunRef5}%
  \BibitemOpen
  \bibfield  {author} {\bibinfo {author} {\bibfnamefont {E.}~\bibnamefont
  {Sicking}}\ and\ \bibinfo {author} {\bibfnamefont {R.}~\bibnamefont
  {Strom}},\ }\bibfield  {title} {\bibinfo {title} {From precision physics to
  the energy frontier with the {Compact Linear Collider}},\ }\href
  {https://doi.org/10.1038/s41567-020-0834-8} {\bibfield  {journal} {\bibinfo
  {journal} {Nature Physics}\ }\textbf {\bibinfo {volume} {16}},\ \bibinfo
  {pages} {386} (\bibinfo {year} {2020})}\BibitemShut {NoStop}%
\bibitem [{\citenamefont {Padamsee}\ \emph {et~al.}(1993)\citenamefont
  {Padamsee}, \citenamefont {Shepard},\ and\ \citenamefont
  {Sundelin}}]{padamsee1993physics}%
  \BibitemOpen
  \bibfield  {author} {\bibinfo {author} {\bibfnamefont {H.}~\bibnamefont
  {Padamsee}}, \bibinfo {author} {\bibfnamefont {K.}~\bibnamefont {Shepard}},\
  and\ \bibinfo {author} {\bibfnamefont {R.}~\bibnamefont {Sundelin}},\
  }\bibfield  {title} {\bibinfo {title} {Physics and accelerator applications
  of rf superconductivity},\ }\href
  {https://doi.org/10.1146/annurev.ns.43.120193.003223} {\bibfield  {journal}
  {\bibinfo  {journal} {Annual Review of Nuclear and Particle Science}\
  }\textbf {\bibinfo {volume} {43}} (\bibinfo {year} {1993})}\BibitemShut
  {NoStop}%
\bibitem [{\citenamefont {Padamsee}(2001)}]{padamsee2001science}%
  \BibitemOpen
  \bibfield  {author} {\bibinfo {author} {\bibfnamefont {H.}~\bibnamefont
  {Padamsee}},\ }\bibfield  {title} {\bibinfo {title} {The science and
  technology of superconducting cavities for accelerators},\ }\href
  {https://doi.org/10.1088/0953-2048/14/4/202} {\bibfield  {journal} {\bibinfo
  {journal} {Superconductor science and technology}\ }\textbf {\bibinfo
  {volume} {14}},\ \bibinfo {pages} {R28} (\bibinfo {year} {2001})}\BibitemShut
  {NoStop}%
\bibitem [{\citenamefont {Posen}\ \emph
  {et~al.}(2015{\natexlab{a}})\citenamefont {Posen}, \citenamefont {Liepe},\
  and\ \citenamefont {Hall}}]{posen2015proof}%
  \BibitemOpen
  \bibfield  {author} {\bibinfo {author} {\bibfnamefont {S.}~\bibnamefont
  {Posen}}, \bibinfo {author} {\bibfnamefont {M.}~\bibnamefont {Liepe}},\ and\
  \bibinfo {author} {\bibfnamefont {D.}~\bibnamefont {Hall}},\ }\bibfield
  {title} {\bibinfo {title} {Proof-of-principle demonstration of
  \protect{Nb$_3$Sn} superconducting radiofrequency cavities for high
  \protect{Q$_0$} applications},\ }\href
  {https://doi.org/https://doi.org/10.1063/1.4913247} {\bibfield  {journal}
  {\bibinfo  {journal} {Applied Physics Letters}\ }\textbf {\bibinfo {volume}
  {106}},\ \bibinfo {pages} {082601} (\bibinfo {year}
  {2015}{\natexlab{a}})}\BibitemShut {NoStop}%
\bibitem [{\citenamefont {Posen}\ and\ \citenamefont
  {Hall}(2017)}]{posen2017nb3sn}%
  \BibitemOpen
  \bibfield  {author} {\bibinfo {author} {\bibfnamefont {S.}~\bibnamefont
  {Posen}}\ and\ \bibinfo {author} {\bibfnamefont {D.~L.}\ \bibnamefont
  {Hall}},\ }\bibfield  {title} {\bibinfo {title} {\protect{Nb$_3$Sn}
  superconducting radiofrequency cavities: fabrication, results, properties,
  and prospects},\ }\href {https://doi.org/10.1088/1361-6668/30/3/033004}
  {\bibfield  {journal} {\bibinfo  {journal} {Superconductor Science and
  Technology}\ }\textbf {\bibinfo {volume} {30}},\ \bibinfo {pages} {033004}
  (\bibinfo {year} {2017})}\BibitemShut {NoStop}%
\bibitem [{\citenamefont {Posen}\ \emph {et~al.}(2021)\citenamefont {Posen},
  \citenamefont {Lee}, \citenamefont {Seidman}, \citenamefont {Romanenko},
  \citenamefont {Tennis}, \citenamefont {Melnychuk},\ and\ \citenamefont
  {Sergatskov}}]{posen2021advances}%
  \BibitemOpen
  \bibfield  {author} {\bibinfo {author} {\bibfnamefont {S.}~\bibnamefont
  {Posen}}, \bibinfo {author} {\bibfnamefont {J.}~\bibnamefont {Lee}}, \bibinfo
  {author} {\bibfnamefont {D.~N.}\ \bibnamefont {Seidman}}, \bibinfo {author}
  {\bibfnamefont {A.}~\bibnamefont {Romanenko}}, \bibinfo {author}
  {\bibfnamefont {B.}~\bibnamefont {Tennis}}, \bibinfo {author} {\bibfnamefont
  {O.}~\bibnamefont {Melnychuk}},\ and\ \bibinfo {author} {\bibfnamefont
  {D.}~\bibnamefont {Sergatskov}},\ }\bibfield  {title} {\bibinfo {title}
  {Advances in \protect{Nb$_3$Sn} superconducting radiofrequency cavities
  towards first practical accelerator applications},\ }\href
  {https://doi.org/10.1088/1361-6668/abc7f7} {\bibfield  {journal} {\bibinfo
  {journal} {Superconductor Science and Technology}\ }\textbf {\bibinfo
  {volume} {34}},\ \bibinfo {pages} {025007} (\bibinfo {year}
  {2021})}\BibitemShut {NoStop}%
\bibitem [{\citenamefont {Burton}\ \emph {et~al.}(2016)\citenamefont {Burton},
  \citenamefont {Beebe}, \citenamefont {Yang}, \citenamefont {Lukaszew},
  \citenamefont {Valente-Feliciano},\ and\ \citenamefont {Reece}}]{Burton}%
  \BibitemOpen
  \bibfield  {author} {\bibinfo {author} {\bibfnamefont {M.~C.}\ \bibnamefont
  {Burton}}, \bibinfo {author} {\bibfnamefont {M.~R.}\ \bibnamefont {Beebe}},
  \bibinfo {author} {\bibfnamefont {K.}~\bibnamefont {Yang}}, \bibinfo {author}
  {\bibfnamefont {R.~A.}\ \bibnamefont {Lukaszew}}, \bibinfo {author}
  {\bibfnamefont {A.-M.}\ \bibnamefont {Valente-Feliciano}},\ and\ \bibinfo
  {author} {\bibfnamefont {C.}~\bibnamefont {Reece}},\ }\bibfield  {title}
  {\bibinfo {title} {Superconducting \protect{NbTiN} thin films for
  superconducting radio frequency accelerator cavity applications},\ }\href
  {https://doi.org/10.1116/1.4941735} {\bibfield  {journal} {\bibinfo
  {journal} {Journal of Vacuum Science \& Technology A}\ }\textbf {\bibinfo
  {volume} {34}},\ \bibinfo {pages} {021518} (\bibinfo {year}
  {2016})}\BibitemShut {NoStop}%
\bibitem [{\citenamefont {Smirnova}\ and\ \citenamefont
  {Starikov}(2017)}]{SunRef6}%
  \BibitemOpen
  \bibfield  {author} {\bibinfo {author} {\bibfnamefont {D.~E.}\ \bibnamefont
  {Smirnova}}\ and\ \bibinfo {author} {\bibfnamefont {S.~V.}\ \bibnamefont
  {Starikov}},\ }\bibfield  {title} {\bibinfo {title} {An interatomic potential
  for simulation of {Zr-Nb} system},\ }\href
  {https://doi.org/10.1016/j.commatsci.2016.12.016} {\bibfield  {journal}
  {\bibinfo  {journal} {Computational Materials Science}\ }\textbf {\bibinfo
  {volume} {129}},\ \bibinfo {pages} {259} (\bibinfo {year}
  {2017})}\BibitemShut {NoStop}%
\bibitem [{\citenamefont {Corsan}\ \emph {et~al.}(1968)\citenamefont {Corsan},
  \citenamefont {Williams}, \citenamefont {Catterall},\ and\ \citenamefont
  {Cook}}]{SunRef1}%
  \BibitemOpen
  \bibfield  {author} {\bibinfo {author} {\bibfnamefont {J.~M.}\ \bibnamefont
  {Corsan}}, \bibinfo {author} {\bibfnamefont {I.}~\bibnamefont {Williams}},
  \bibinfo {author} {\bibfnamefont {J.~A.}\ \bibnamefont {Catterall}},\ and\
  \bibinfo {author} {\bibfnamefont {A.~J.}\ \bibnamefont {Cook}},\ }\bibfield
  {title} {\bibinfo {title} {Superconductor transition temperatures of
  zirconium-niobium alloys},\ }\href
  {https://doi.org/10.1016/0022-5088(68)90109-4} {\bibfield  {journal}
  {\bibinfo  {journal} {Journal of the Less-Common Metals}\ }\textbf {\bibinfo
  {volume} {15}},\ \bibinfo {pages} {437} (\bibinfo {year} {1968})}\BibitemShut
  {NoStop}%
\bibitem [{\citenamefont {Hulm}\ and\ \citenamefont
  {Blaugher}(1961)}]{hulm1961}%
  \BibitemOpen
  \bibfield  {author} {\bibinfo {author} {\bibfnamefont {J.~K.}\ \bibnamefont
  {Hulm}}\ and\ \bibinfo {author} {\bibfnamefont {R.~D.}\ \bibnamefont
  {Blaugher}},\ }\bibfield  {title} {\bibinfo {title} {Superconducting solid
  solution alloys of the transition elements},\ }\href
  {https://doi.org/10.1103/PhysRev.123.1569} {\bibfield  {journal} {\bibinfo
  {journal} {Physical Review}\ }\textbf {\bibinfo {volume} {123}},\ \bibinfo
  {pages} {1569} (\bibinfo {year} {1961})}\BibitemShut {NoStop}%
\bibitem [{\citenamefont {Chapman}(1995)}]{chapman1995superheating}%
  \BibitemOpen
  \bibfield  {author} {\bibinfo {author} {\bibfnamefont {S.~J.}\ \bibnamefont
  {Chapman}},\ }\bibfield  {title} {\bibinfo {title} {Superheating field of
  type {II} superconductors},\ }\href {https://www.jstor.org/stable/2102573}
  {\bibfield  {journal} {\bibinfo  {journal} {SIAM Journal on Applied
  Mathematics}\ }\textbf {\bibinfo {volume} {55}},\ \bibinfo {pages} {1233}
  (\bibinfo {year} {1995})}\BibitemShut {NoStop}%
\bibitem [{\citenamefont {Dolgert}\ \emph {et~al.}(1996)\citenamefont
  {Dolgert}, \citenamefont {Di~Bartolo},\ and\ \citenamefont
  {Dorsey}}]{dolgert1996superheating}%
  \BibitemOpen
  \bibfield  {author} {\bibinfo {author} {\bibfnamefont {A.~J.}\ \bibnamefont
  {Dolgert}}, \bibinfo {author} {\bibfnamefont {S.~J.}\ \bibnamefont
  {Di~Bartolo}},\ and\ \bibinfo {author} {\bibfnamefont {A.~T.}\ \bibnamefont
  {Dorsey}},\ }\bibfield  {title} {\bibinfo {title} {Superheating fields of
  superconductors: Asymptotic analysis and numerical results},\ }\href
  {https://doi.org/10.1103/PhysRevB.53.5650} {\bibfield  {journal} {\bibinfo
  {journal} {Physical Review B}\ }\textbf {\bibinfo {volume} {53}},\ \bibinfo
  {pages} {5650} (\bibinfo {year} {1996})}\BibitemShut {NoStop}%
\bibitem [{\citenamefont {Transtrum}\ \emph {et~al.}(2011)\citenamefont
  {Transtrum}, \citenamefont {Catelani},\ and\ \citenamefont
  {Sethna}}]{transtrum2011superheating}%
  \BibitemOpen
  \bibfield  {author} {\bibinfo {author} {\bibfnamefont {M.~K.}\ \bibnamefont
  {Transtrum}}, \bibinfo {author} {\bibfnamefont {G.}~\bibnamefont
  {Catelani}},\ and\ \bibinfo {author} {\bibfnamefont {J.~P.}\ \bibnamefont
  {Sethna}},\ }\bibfield  {title} {\bibinfo {title} {Superheating field of
  superconductors within {Ginzburg-Landau} theory},\ }\href
  {https://doi.org/10.1103/PhysRevB.83.094505} {\bibfield  {journal} {\bibinfo
  {journal} {Physical Review B}\ }\textbf {\bibinfo {volume} {83}},\ \bibinfo
  {pages} {094505} (\bibinfo {year} {2011})}\BibitemShut {NoStop}%
\bibitem [{\citenamefont {Catelani}\ and\ \citenamefont
  {Sethna}(2008)}]{catelani2008temperature}%
  \BibitemOpen
  \bibfield  {author} {\bibinfo {author} {\bibfnamefont {G.}~\bibnamefont
  {Catelani}}\ and\ \bibinfo {author} {\bibfnamefont {J.~P.}\ \bibnamefont
  {Sethna}},\ }\bibfield  {title} {\bibinfo {title} {Temperature dependence of
  the superheating field for superconductors in the high-$\kappa$ {London}
  limit},\ }\href {https://doi.org/10.1103/PhysRevB.78.224509} {\bibfield
  {journal} {\bibinfo  {journal} {Physical Review B}\ }\textbf {\bibinfo
  {volume} {78}},\ \bibinfo {pages} {224509} (\bibinfo {year}
  {2008})}\BibitemShut {NoStop}%
\bibitem [{\citenamefont {Lin}\ and\ \citenamefont
  {Gurevich}(2012)}]{lin2012effect}%
  \BibitemOpen
  \bibfield  {author} {\bibinfo {author} {\bibfnamefont {F.~P.-J.}\
  \bibnamefont {Lin}}\ and\ \bibinfo {author} {\bibfnamefont {A.}~\bibnamefont
  {Gurevich}},\ }\bibfield  {title} {\bibinfo {title} {Effect of impurities on
  the superheating field of type-\protect{II} superconductors},\ }\href
  {https://doi.org/10.1103/PhysRevB.85.054513} {\bibfield  {journal} {\bibinfo
  {journal} {Physical Review B}\ }\textbf {\bibinfo {volume} {85}},\ \bibinfo
  {pages} {054513} (\bibinfo {year} {2012})}\BibitemShut {NoStop}%
\bibitem [{\citenamefont {Liarte}\ \emph {et~al.}(2016)\citenamefont {Liarte},
  \citenamefont {Transtrum},\ and\ \citenamefont
  {Sethna}}]{liarte2016ginzburg}%
  \BibitemOpen
  \bibfield  {author} {\bibinfo {author} {\bibfnamefont {D.~B.}\ \bibnamefont
  {Liarte}}, \bibinfo {author} {\bibfnamefont {M.~K.}\ \bibnamefont
  {Transtrum}},\ and\ \bibinfo {author} {\bibfnamefont {J.~P.}\ \bibnamefont
  {Sethna}},\ }\bibfield  {title} {\bibinfo {title} {Ginzburg-landau theory of
  the superheating field anisotropy of layered superconductors},\ }\href
  {https://doi.org/10.1103/PhysRevB.94.144504} {\bibfield  {journal} {\bibinfo
  {journal} {Physical Review B}\ }\textbf {\bibinfo {volume} {94}},\ \bibinfo
  {pages} {144504} (\bibinfo {year} {2016})}\BibitemShut {NoStop}%
\bibitem [{\citenamefont {Kubo}\ \emph {et~al.}(2014)\citenamefont {Kubo},
  \citenamefont {Iwashita},\ and\ \citenamefont {Saeki}}]{kubo2014radio}%
  \BibitemOpen
  \bibfield  {author} {\bibinfo {author} {\bibfnamefont {T.}~\bibnamefont
  {Kubo}}, \bibinfo {author} {\bibfnamefont {Y.}~\bibnamefont {Iwashita}},\
  and\ \bibinfo {author} {\bibfnamefont {T.}~\bibnamefont {Saeki}},\ }\bibfield
   {title} {\bibinfo {title} {Radio-frequency electromagnetic field and vortex
  penetration in multilayered superconductors},\ }\href
  {https://doi.org/10.1063/1.4862892} {\bibfield  {journal} {\bibinfo
  {journal} {Applied Physics Letters}\ }\textbf {\bibinfo {volume} {104}},\
  \bibinfo {pages} {032603} (\bibinfo {year} {2014})}\BibitemShut {NoStop}%
\bibitem [{\citenamefont {Gurevich}(2015)}]{gurevich2015maximum}%
  \BibitemOpen
  \bibfield  {author} {\bibinfo {author} {\bibfnamefont {A.}~\bibnamefont
  {Gurevich}},\ }\bibfield  {title} {\bibinfo {title} {Maximum screening fields
  of superconducting multilayer structures},\ }\href
  {https://doi.org/10.1063/1.4905711} {\bibfield  {journal} {\bibinfo
  {journal} {AIP Advances}\ }\textbf {\bibinfo {volume} {5}},\ \bibinfo {pages}
  {017112} (\bibinfo {year} {2015})}\BibitemShut {NoStop}%
\bibitem [{\citenamefont {Posen}\ \emph
  {et~al.}(2015{\natexlab{b}})\citenamefont {Posen}, \citenamefont {Transtrum},
  \citenamefont {Catelani}, \citenamefont {Liepe},\ and\ \citenamefont
  {Sethna}}]{posen2015shielding}%
  \BibitemOpen
  \bibfield  {author} {\bibinfo {author} {\bibfnamefont {S.}~\bibnamefont
  {Posen}}, \bibinfo {author} {\bibfnamefont {M.~K.}\ \bibnamefont
  {Transtrum}}, \bibinfo {author} {\bibfnamefont {G.}~\bibnamefont {Catelani}},
  \bibinfo {author} {\bibfnamefont {M.~U.}\ \bibnamefont {Liepe}},\ and\
  \bibinfo {author} {\bibfnamefont {J.~P.}\ \bibnamefont {Sethna}},\ }\bibfield
   {title} {\bibinfo {title} {Shielding superconductors with thin films as
  applied to rf cavities for particle accelerators},\ }\href
  {https://doi.org/10.1103/PhysRevApplied.4.044019} {\bibfield  {journal}
  {\bibinfo  {journal} {Physical Review Applied}\ }\textbf {\bibinfo {volume}
  {4}},\ \bibinfo {pages} {044019} (\bibinfo {year}
  {2015}{\natexlab{b}})}\BibitemShut {NoStop}%
\bibitem [{\citenamefont {Kubo}(2016)}]{kubo2016multilayer}%
  \BibitemOpen
  \bibfield  {author} {\bibinfo {author} {\bibfnamefont {T.}~\bibnamefont
  {Kubo}},\ }\bibfield  {title} {\bibinfo {title} {Multilayer coating for
  higher accelerating fields in superconducting radio-frequency cavities: a
  review of theoretical aspects},\ }\href
  {https://doi.org/10.1088/1361-6668/30/2/023001} {\bibfield  {journal}
  {\bibinfo  {journal} {Superconductor Science and Technology}\ }\textbf
  {\bibinfo {volume} {30}},\ \bibinfo {pages} {023001} (\bibinfo {year}
  {2016})}\BibitemShut {NoStop}%
\bibitem [{\citenamefont {Ngampruetikorn}\ and\ \citenamefont
  {Sauls}(2019)}]{ngampruetikorn2019effect}%
  \BibitemOpen
  \bibfield  {author} {\bibinfo {author} {\bibfnamefont {V.}~\bibnamefont
  {Ngampruetikorn}}\ and\ \bibinfo {author} {\bibfnamefont {J.}~\bibnamefont
  {Sauls}},\ }\bibfield  {title} {\bibinfo {title} {Effect of inhomogeneous
  surface disorder on the superheating field of superconducting rf cavities},\
  }\href {https://doi.org/10.1103/PhysRevResearch.1.012015} {\bibfield
  {journal} {\bibinfo  {journal} {Physical Review Research}\ }\textbf {\bibinfo
  {volume} {1}},\ \bibinfo {pages} {012015} (\bibinfo {year}
  {2019})}\BibitemShut {NoStop}%
\bibitem [{\citenamefont {Pack}\ \emph {et~al.}(2020)\citenamefont {Pack},
  \citenamefont {Carlson}, \citenamefont {Wadsworth},\ and\ \citenamefont
  {Transtrum}}]{pack2020vortex}%
  \BibitemOpen
  \bibfield  {author} {\bibinfo {author} {\bibfnamefont {A.~R.}\ \bibnamefont
  {Pack}}, \bibinfo {author} {\bibfnamefont {J.}~\bibnamefont {Carlson}},
  \bibinfo {author} {\bibfnamefont {S.}~\bibnamefont {Wadsworth}},\ and\
  \bibinfo {author} {\bibfnamefont {M.~K.}\ \bibnamefont {Transtrum}},\
  }\bibfield  {title} {\bibinfo {title} {Vortex nucleation in superconductors
  within time-dependent ginzburg-landau theory in two and three dimensions:
  role of surface defects and material inhomogeneities},\ }\href
  {https://doi.org/10.1103/PhysRevB.101.144504} {\bibfield  {journal} {\bibinfo
   {journal} {Physical Review B}\ }\textbf {\bibinfo {volume} {101}},\ \bibinfo
  {pages} {144504} (\bibinfo {year} {2020})}\BibitemShut {NoStop}%
\bibitem [{\citenamefont {Giannozzi}\ \emph {et~al.}(2009)\citenamefont
  {Giannozzi} \emph {et~al.}}]{qe1}%
  \BibitemOpen
  \bibfield  {author} {\bibinfo {author} {\bibfnamefont {P.}~\bibnamefont
  {Giannozzi}} \emph {et~al.},\ }\bibfield  {title} {\bibinfo {title}
  {{QUANTUM} {ESPRESSO}: a modular and open-source software project for quantum
  simulations of materials},\ }\href
  {https://doi.org/10.1088/0953-8984/21/39/395502} {\bibfield  {journal}
  {\bibinfo  {journal} {Journal of Physics: Condensed Matter}\ }\textbf
  {\bibinfo {volume} {21}},\ \bibinfo {pages} {395502} (\bibinfo {year}
  {2009})}\BibitemShut {NoStop}%
\bibitem [{\citenamefont {Giannozzi}\ \emph {et~al.}(2020)\citenamefont
  {Giannozzi} \emph {et~al.}}]{qe2}%
  \BibitemOpen
  \bibfield  {author} {\bibinfo {author} {\bibfnamefont {P.}~\bibnamefont
  {Giannozzi}} \emph {et~al.},\ }\bibfield  {title} {\bibinfo {title} {Quantum
  {ESPRESSO} toward the exascale},\ }\href {https://doi.org/10.1063/5.0005082}
  {\bibfield  {journal} {\bibinfo  {journal} {The Journal of Chemical Physics}\
  }\textbf {\bibinfo {volume} {152}},\ \bibinfo {pages} {154105} (\bibinfo
  {year} {2020})}\BibitemShut {NoStop}%
\bibitem [{\citenamefont {Giannozzi}\ \emph {et~al.}(2017)\citenamefont
  {Giannozzi} \emph {et~al.}}]{qe3}%
  \BibitemOpen
  \bibfield  {author} {\bibinfo {author} {\bibfnamefont {P.}~\bibnamefont
  {Giannozzi}} \emph {et~al.},\ }\bibfield  {title} {\bibinfo {title} {Advanced
  capabilities for materials modelling with quantum {ESPRESSO}},\ }\href
  {https://doi.org/10.1088/1361-648x/aa8f79} {\bibfield  {journal} {\bibinfo
  {journal} {Journal of Physics: Condensed Matter}\ }\textbf {\bibinfo {volume}
  {29}},\ \bibinfo {pages} {465901} (\bibinfo {year} {2017})}\BibitemShut
  {NoStop}%
\bibitem [{\citenamefont {Perdew}\ \emph {et~al.}(1996)\citenamefont {Perdew},
  \citenamefont {Burke},\ and\ \citenamefont {Ernzerhof}}]{PBE}%
  \BibitemOpen
  \bibfield  {author} {\bibinfo {author} {\bibfnamefont {J.~P.}\ \bibnamefont
  {Perdew}}, \bibinfo {author} {\bibfnamefont {K.}~\bibnamefont {Burke}},\ and\
  \bibinfo {author} {\bibfnamefont {M.}~\bibnamefont {Ernzerhof}},\ }\bibfield
  {title} {\bibinfo {title} {Generalized gradient approximation made simple},\
  }\href {https://doi.org/10.1103/PhysRevLett.77.3865} {\bibfield  {journal}
  {\bibinfo  {journal} {Physical Review Letter}\ }\textbf {\bibinfo {volume}
  {77}},\ \bibinfo {pages} {3865} (\bibinfo {year} {1996})}\BibitemShut
  {NoStop}%
\bibitem [{\citenamefont {Hamann}(2013)}]{Hamann2013}%
  \BibitemOpen
  \bibfield  {author} {\bibinfo {author} {\bibfnamefont {D.~R.}\ \bibnamefont
  {Hamann}},\ }\bibfield  {title} {\bibinfo {title} {Optimized norm-conserving
  {Vanderbilt} pseudopotentials},\ }\href
  {https://doi.org/10.1103/PhysRevB.88.085117} {\bibfield  {journal} {\bibinfo
  {journal} {Physical Review B}\ }\textbf {\bibinfo {volume} {95}},\ \bibinfo
  {pages} {239906} (\bibinfo {year} {2013})}\BibitemShut {NoStop}%
\bibitem [{\citenamefont {Schlipf}\ and\ \citenamefont
  {Gygi}(2015)}]{Schlipf2015}%
  \BibitemOpen
  \bibfield  {author} {\bibinfo {author} {\bibfnamefont {M.}~\bibnamefont
  {Schlipf}}\ and\ \bibinfo {author} {\bibfnamefont {F.}~\bibnamefont {Gygi}},\
  }\bibfield  {title} {\bibinfo {title} {Optimization algorithm for the
  generation of {ONCV} pseudopotentials},\ }\href
  {https://doi.org/10.1016/j.cpc.2015.05.011} {\bibfield  {journal} {\bibinfo
  {journal} {Computer Physics Communications}\ }\textbf {\bibinfo {volume}
  {196}},\ \bibinfo {pages} {36} (\bibinfo {year} {2015})}\BibitemShut
  {NoStop}%
\bibitem [{\citenamefont {Methfessel}\ and\ \citenamefont
  {Paxton}(1989)}]{Methfessel1989}%
  \BibitemOpen
  \bibfield  {author} {\bibinfo {author} {\bibfnamefont {M.}~\bibnamefont
  {Methfessel}}\ and\ \bibinfo {author} {\bibfnamefont {A.~T.}\ \bibnamefont
  {Paxton}},\ }\bibfield  {title} {\bibinfo {title} {High-precision sampling
  for {Brillouin}-zone integration in metals},\ }\href
  {https://doi.org/10.1103/physrevb.40.3616} {\bibfield  {journal} {\bibinfo
  {journal} {Physical Review B}\ }\textbf {\bibinfo {volume} {40}},\ \bibinfo
  {pages} {3616} (\bibinfo {year} {1989})}\BibitemShut {NoStop}%
\bibitem [{\citenamefont {Kawamura}\ \emph {et~al.}(2014)\citenamefont
  {Kawamura}, \citenamefont {Gohda},\ and\ \citenamefont
  {Tsuneyuki}}]{PhysRevB.89.094515}%
  \BibitemOpen
  \bibfield  {author} {\bibinfo {author} {\bibfnamefont {M.}~\bibnamefont
  {Kawamura}}, \bibinfo {author} {\bibfnamefont {Y.}~\bibnamefont {Gohda}},\
  and\ \bibinfo {author} {\bibfnamefont {S.}~\bibnamefont {Tsuneyuki}},\
  }\bibfield  {title} {\bibinfo {title} {Improved tetrahedron method for the
  {Brillouin}-zone integration applicable to response functions},\ }\href
  {https://doi.org/10.1103/PhysRevB.89.094515} {\bibfield  {journal} {\bibinfo
  {journal} {Physical Review B}\ }\textbf {\bibinfo {volume} {89}},\ \bibinfo
  {pages} {094515} (\bibinfo {year} {2014})}\BibitemShut {NoStop}%
\bibitem [{\citenamefont {Giustino}\ \emph {et~al.}(2007)\citenamefont
  {Giustino}, \citenamefont {Cohen},\ and\ \citenamefont {Louie}}]{epw1}%
  \BibitemOpen
  \bibfield  {author} {\bibinfo {author} {\bibfnamefont {F.}~\bibnamefont
  {Giustino}}, \bibinfo {author} {\bibfnamefont {M.~L.}\ \bibnamefont
  {Cohen}},\ and\ \bibinfo {author} {\bibfnamefont {S.~G.}\ \bibnamefont
  {Louie}},\ }\bibfield  {title} {\bibinfo {title} {Electron-phonon interaction
  using {Wannier} functions},\ }\href
  {https://doi.org/10.1103/PhysRevB.76.165108} {\bibfield  {journal} {\bibinfo
  {journal} {Physical Review B}\ }\textbf {\bibinfo {volume} {76}} (\bibinfo
  {year} {2007})}\BibitemShut {NoStop}%
\bibitem [{\citenamefont {Ponc{\'{e}}}\ \emph {et~al.}(2016)\citenamefont
  {Ponc{\'{e}}}, \citenamefont {Margine}, \citenamefont {Verdi},\ and\
  \citenamefont {Giustino}}]{epw2}%
  \BibitemOpen
  \bibfield  {author} {\bibinfo {author} {\bibfnamefont {S.}~\bibnamefont
  {Ponc{\'{e}}}}, \bibinfo {author} {\bibfnamefont {E.}~\bibnamefont
  {Margine}}, \bibinfo {author} {\bibfnamefont {C.}~\bibnamefont {Verdi}},\
  and\ \bibinfo {author} {\bibfnamefont {F.}~\bibnamefont {Giustino}},\
  }\bibfield  {title} {\bibinfo {title} {{EPW}: Electron{\textendash}phonon
  coupling, transport and superconducting properties using maximally localized
  {Wannier} functions},\ }\href {https://doi.org/10.1016/j.cpc.2016.07.028}
  {\bibfield  {journal} {\bibinfo  {journal} {Computer Physics Communications}\
  }\textbf {\bibinfo {volume} {209}},\ \bibinfo {pages} {116} (\bibinfo {year}
  {2016})}\BibitemShut {NoStop}%
\bibitem [{\citenamefont {Allen}\ and\ \citenamefont
  {Dynes}(1975)}]{allen-dynes}%
  \BibitemOpen
  \bibfield  {author} {\bibinfo {author} {\bibfnamefont {P.~B.}\ \bibnamefont
  {Allen}}\ and\ \bibinfo {author} {\bibfnamefont {R.~C.}\ \bibnamefont
  {Dynes}},\ }\bibfield  {title} {\bibinfo {title} {Transition temperature of
  strong-coupled superconductors reanalyzed},\ }\href
  {https://doi.org/10.1103/PhysRevB.12.905} {\bibfield  {journal} {\bibinfo
  {journal} {Physical Review B}\ }\textbf {\bibinfo {volume} {12}},\ \bibinfo
  {pages} {905} (\bibinfo {year} {1975})}\BibitemShut {NoStop}%
\bibitem [{\citenamefont {Xie}\ \emph {et~al.}(2022)\citenamefont {Xie},
  \citenamefont {Quan}, \citenamefont {Hire}, \citenamefont {Deng},
  \citenamefont {DeStefano}, \citenamefont {Salinas}, \citenamefont {Shah},
  \citenamefont {Fanfarillo}, \citenamefont {Lim}, \citenamefont {Kim},
  \citenamefont {Stewart}, \citenamefont {Hamlin}, \citenamefont {Hirschfeld},\
  and\ \citenamefont {Hennig}}]{Xie_2022}%
  \BibitemOpen
  \bibfield  {author} {\bibinfo {author} {\bibfnamefont {S.~R.}\ \bibnamefont
  {Xie}}, \bibinfo {author} {\bibfnamefont {Y.}~\bibnamefont {Quan}}, \bibinfo
  {author} {\bibfnamefont {A.~C.}\ \bibnamefont {Hire}}, \bibinfo {author}
  {\bibfnamefont {B.}~\bibnamefont {Deng}}, \bibinfo {author} {\bibfnamefont
  {J.~M.}\ \bibnamefont {DeStefano}}, \bibinfo {author} {\bibfnamefont
  {I.}~\bibnamefont {Salinas}}, \bibinfo {author} {\bibfnamefont {U.~S.}\
  \bibnamefont {Shah}}, \bibinfo {author} {\bibfnamefont {L.}~\bibnamefont
  {Fanfarillo}}, \bibinfo {author} {\bibfnamefont {J.}~\bibnamefont {Lim}},
  \bibinfo {author} {\bibfnamefont {J.}~\bibnamefont {Kim}}, \bibinfo {author}
  {\bibfnamefont {G.~R.}\ \bibnamefont {Stewart}}, \bibinfo {author}
  {\bibfnamefont {J.~J.}\ \bibnamefont {Hamlin}}, \bibinfo {author}
  {\bibfnamefont {P.~J.}\ \bibnamefont {Hirschfeld}},\ and\ \bibinfo {author}
  {\bibfnamefont {R.~G.}\ \bibnamefont {Hennig}},\ }\bibfield  {title}
  {\bibinfo {title} {Machine learning of superconducting critical temperature
  from eliashberg theory},\ }\href {https://doi.org/10.1038/s41524-021-00666-7}
  {\bibfield  {journal} {\bibinfo  {journal} {npj Computational Materials}\
  }\textbf {\bibinfo {volume} {8}} (\bibinfo {year} {2022})}\BibitemShut
  {NoStop}%
\bibitem [{\citenamefont {Bardeen}\ \emph {et~al.}(1957)\citenamefont
  {Bardeen}, \citenamefont {Cooper},\ and\ \citenamefont {Schrieffer}}]{BCS}%
  \BibitemOpen
  \bibfield  {author} {\bibinfo {author} {\bibfnamefont {J.}~\bibnamefont
  {Bardeen}}, \bibinfo {author} {\bibfnamefont {L.~N.}\ \bibnamefont
  {Cooper}},\ and\ \bibinfo {author} {\bibfnamefont {J.~R.}\ \bibnamefont
  {Schrieffer}},\ }\bibfield  {title} {\bibinfo {title} {Theory of
  superconductivity},\ }\href {https://doi.org/10.1103/PhysRev.108.1175}
  {\bibfield  {journal} {\bibinfo  {journal} {Physical Review}\ }\textbf
  {\bibinfo {volume} {108}},\ \bibinfo {pages} {1175} (\bibinfo {year}
  {1957})}\BibitemShut {NoStop}%
\bibitem [{\citenamefont {Sundararaman}\ \emph {et~al.}(2017)\citenamefont
  {Sundararaman}, \citenamefont {Letchworth-Weaver}, \citenamefont {Schwarz},
  \citenamefont {Gunceler}, \citenamefont {Ozhabes},\ and\ \citenamefont
  {Arias}}]{Sundararaman}%
  \BibitemOpen
  \bibfield  {author} {\bibinfo {author} {\bibfnamefont {R.}~\bibnamefont
  {Sundararaman}}, \bibinfo {author} {\bibfnamefont {K.}~\bibnamefont
  {Letchworth-Weaver}}, \bibinfo {author} {\bibfnamefont {K.~A.}\ \bibnamefont
  {Schwarz}}, \bibinfo {author} {\bibfnamefont {D.}~\bibnamefont {Gunceler}},
  \bibinfo {author} {\bibfnamefont {Y.}~\bibnamefont {Ozhabes}},\ and\ \bibinfo
  {author} {\bibfnamefont {T.~A.}\ \bibnamefont {Arias}},\ }\bibfield  {title}
  {\bibinfo {title} {{JDFTx}: Software for joint density-functional theory},\
  }\href {https://doi.org/10.1016/j.softx.2017.10.006} {\bibfield  {journal}
  {\bibinfo  {journal} {SoftwareX}\ }\textbf {\bibinfo {volume} {6}},\ \bibinfo
  {pages} {278} (\bibinfo {year} {2017})}\BibitemShut {NoStop}%
\bibitem [{\citenamefont {Garrity}\ \emph {et~al.}(2014)\citenamefont
  {Garrity}, \citenamefont {Bennett}, \citenamefont {Rabe},\ and\ \citenamefont
  {Vanderbilt}}]{Psp}%
  \BibitemOpen
  \bibfield  {author} {\bibinfo {author} {\bibfnamefont {K.~F.}\ \bibnamefont
  {Garrity}}, \bibinfo {author} {\bibfnamefont {J.~W.}\ \bibnamefont
  {Bennett}}, \bibinfo {author} {\bibfnamefont {K.~M.}\ \bibnamefont {Rabe}},\
  and\ \bibinfo {author} {\bibfnamefont {D.}~\bibnamefont {Vanderbilt}},\
  }\bibfield  {title} {\bibinfo {title} {Pseudopotentials for high-throughput
  {DFT} calculations},\ }\href
  {https://doi.org/10.1016/j.commatsci.2013.08.053} {\bibfield  {journal}
  {\bibinfo  {journal} {Computational Materials Science}\ }\textbf {\bibinfo
  {volume} {81}},\ \bibinfo {pages} {446} (\bibinfo {year} {2014})}\BibitemShut
  {NoStop}%
\bibitem [{\citenamefont {Feng}\ and\ \citenamefont {Widom}(2018)}]{widom2018}%
  \BibitemOpen
  \bibfield  {author} {\bibinfo {author} {\bibfnamefont {B.}~\bibnamefont
  {Feng}}\ and\ \bibinfo {author} {\bibfnamefont {M.}~\bibnamefont {Widom}},\
  }\bibfield  {title} {\bibinfo {title} {Band structure theory of the bcc to
  hcp {Burgers} distortion},\ }\href
  {https://doi.org/10.1103/PhysRevB.98.174108} {\bibfield  {journal} {\bibinfo
  {journal} {Physical Review B}\ }\textbf {\bibinfo {volume} {98}},\ \bibinfo
  {pages} {174108} (\bibinfo {year} {2018})}\BibitemShut {NoStop}%
\bibitem [{\citenamefont {Bekaert}\ \emph {et~al.}(2018)\citenamefont
  {Bekaert}, \citenamefont {Aperis}, \citenamefont {Partoens}, \citenamefont
  {Oppeneer},\ and\ \citenamefont {Milo\ifmmode \check{s}\else
  \v{s}\fi{}evi\ifmmode~\acute{c}\else \'{c}\fi{}}}]{Bekaert}%
  \BibitemOpen
  \bibfield  {author} {\bibinfo {author} {\bibfnamefont {J.}~\bibnamefont
  {Bekaert}}, \bibinfo {author} {\bibfnamefont {A.}~\bibnamefont {Aperis}},
  \bibinfo {author} {\bibfnamefont {B.}~\bibnamefont {Partoens}}, \bibinfo
  {author} {\bibfnamefont {P.~M.}\ \bibnamefont {Oppeneer}},\ and\ \bibinfo
  {author} {\bibfnamefont {M.~V.}\ \bibnamefont {Milo\ifmmode \check{s}\else
  \v{s}\fi{}evi\ifmmode~\acute{c}\else \'{c}\fi{}}},\ }\bibfield  {title}
  {\bibinfo {title} {Advanced first-principles theory of superconductivity
  including both lattice vibrations and spin fluctuations: The case of
  {${\mathrm{FeB}}_{4}$}},\ }\href {https://doi.org/10.1103/PhysRevB.97.014503}
  {\bibfield  {journal} {\bibinfo  {journal} {Physical Review B}\ }\textbf
  {\bibinfo {volume} {97}},\ \bibinfo {pages} {014503} (\bibinfo {year}
  {2018})}\BibitemShut {NoStop}%
\bibitem [{\citenamefont {Thompson}\ \emph {et~al.}(2003)\citenamefont
  {Thompson}, \citenamefont {Banerjee}, \citenamefont {Dregia},\ and\
  \citenamefont {Fraser}}]{SunRef8}%
  \BibitemOpen
  \bibfield  {author} {\bibinfo {author} {\bibfnamefont {G.~B.}\ \bibnamefont
  {Thompson}}, \bibinfo {author} {\bibfnamefont {R.}~\bibnamefont {Banerjee}},
  \bibinfo {author} {\bibfnamefont {S.~A.}\ \bibnamefont {Dregia}},\ and\
  \bibinfo {author} {\bibfnamefont {H.~L.}\ \bibnamefont {Fraser}},\ }\bibfield
   {title} {\bibinfo {title} {Phase stability of bcc {Zr} in {Nb/Zr} thin film
  multilayers},\ }\href {https://doi.org/10.1016/S1359-6454(03)00380-X}
  {\bibfield  {journal} {\bibinfo  {journal} {Acta Materialia}\ }\textbf
  {\bibinfo {volume} {51}},\ \bibinfo {pages} {5285} (\bibinfo {year}
  {2003})}\BibitemShut {NoStop}%
\bibitem [{\citenamefont {Wang}\ \emph {et~al.}(2011)\citenamefont {Wang},
  \citenamefont {Zhang}, \citenamefont {Liu}, \citenamefont {Li},\ and\
  \citenamefont {Zhang}}]{SunRef7}%
  \BibitemOpen
  \bibfield  {author} {\bibinfo {author} {\bibfnamefont {B.~T.}\ \bibnamefont
  {Wang}}, \bibinfo {author} {\bibfnamefont {P.}~\bibnamefont {Zhang}},
  \bibinfo {author} {\bibfnamefont {H.~Y.}\ \bibnamefont {Liu}}, \bibinfo
  {author} {\bibfnamefont {W.~D.}\ \bibnamefont {Li}},\ and\ \bibinfo {author}
  {\bibfnamefont {P.}~\bibnamefont {Zhang}},\ }\bibfield  {title} {\bibinfo
  {title} {First-principles calculations of phase transition, low elastic
  modulus, and superconductivity for zirconium},\ }\href
  {https://doi.org/10.1063/1.3556753} {\bibfield  {journal} {\bibinfo
  {journal} {Journal of Applied Physics}\ }\textbf {\bibinfo {volume} {109}},\
  \bibinfo {pages} {063514} (\bibinfo {year} {2011})}\BibitemShut {NoStop}%
\bibitem [{\citenamefont {Cavalleri}\ \emph {et~al.}(1989)\citenamefont
  {Cavalleri}, \citenamefont {Giacomozzi}, \citenamefont {Guzman},\ and\
  \citenamefont {Ossi}}]{SunRef2}%
  \BibitemOpen
  \bibfield  {author} {\bibinfo {author} {\bibfnamefont {A.}~\bibnamefont
  {Cavalleri}}, \bibinfo {author} {\bibfnamefont {F.}~\bibnamefont
  {Giacomozzi}}, \bibinfo {author} {\bibfnamefont {L.}~\bibnamefont {Guzman}},\
  and\ \bibinfo {author} {\bibfnamefont {P.~M.}\ \bibnamefont {Ossi}},\
  }\bibfield  {title} {\bibinfo {title} {Structure and superconductivity of
  {Nb-Zr} thin films},\ }\href {https://doi.org/10.1088/0953-8984/1/37/015}
  {\bibfield  {journal} {\bibinfo  {journal} {Journal of Physics: Condensed
  Matter}\ }\textbf {\bibinfo {volume} {1}},\ \bibinfo {pages} {6685} (\bibinfo
  {year} {1989})}\BibitemShut {NoStop}%
\bibitem [{\citenamefont {Oseroff}\ \emph {et~al.}(2021)\citenamefont
  {Oseroff}, \citenamefont {Liepe},\ and\ \citenamefont {Sun}}]{SunRef3}%
  \BibitemOpen
  \bibfield  {author} {\bibinfo {author} {\bibfnamefont {T.}~\bibnamefont
  {Oseroff}}, \bibinfo {author} {\bibfnamefont {M.}~\bibnamefont {Liepe}},\
  and\ \bibinfo {author} {\bibfnamefont {Z.}~\bibnamefont {Sun}},\ }\bibfield
  {title} {\bibinfo {title} {Sample test systems for next-gen {SRF} surfaces},\
  }in\ \href {https://doi.org/10.18429/JACoW-SRF2021-TU0FDV07} {\emph {\bibinfo
  {booktitle} {Proceedings of 20th International Conference on RF
  Superconductivity}}}\ (\bibinfo  {publisher} {JACoW Publishing},\ \bibinfo
  {address} {East Lansing, MI, USA},\ \bibinfo {year} {2021})\ p.\ \bibinfo
  {pages} {357}\BibitemShut {NoStop}%
\bibitem [{\citenamefont {Tinkham}(2004)}]{tinkham2004introduction}%
  \BibitemOpen
  \bibfield  {author} {\bibinfo {author} {\bibfnamefont {M.}~\bibnamefont
  {Tinkham}},\ }\href@noop {} {\emph {\bibinfo {title} {Introduction to
  superconductivity}}}\ (\bibinfo  {publisher} {Courier Corporation},\ \bibinfo
  {year} {2004})\BibitemShut {NoStop}%
\bibitem [{\citenamefont {Kopnin}(2001)}]{kopnin2001theory}%
  \BibitemOpen
  \bibfield  {author} {\bibinfo {author} {\bibfnamefont {N.}~\bibnamefont
  {Kopnin}},\ }\href
  {https://doi.org/10.1093/acprof:oso/9780198507888.001.0001} {\emph {\bibinfo
  {title} {Theory of nonequilibrium superconductivity}}},\ Vol.\ \bibinfo
  {volume} {110}\ (\bibinfo  {publisher} {Oxford University Press},\ \bibinfo
  {year} {2001})\BibitemShut {NoStop}%
\bibitem [{\citenamefont {De~Gennes}\ and\ \citenamefont
  {Pincus}(2018)}]{de2018superconductivity}%
  \BibitemOpen
  \bibfield  {author} {\bibinfo {author} {\bibfnamefont {P.-G.}\ \bibnamefont
  {De~Gennes}}\ and\ \bibinfo {author} {\bibfnamefont {P.~A.}\ \bibnamefont
  {Pincus}},\ }\href {https://doi.org/10.1201/9780429497032} {\emph {\bibinfo
  {title} {Superconductivity of metals and alloys}}}\ (\bibinfo  {publisher}
  {CRC Press},\ \bibinfo {year} {2018})\BibitemShut {NoStop}%
\end{thebibliography}
\end{document}